\documentclass[twocolumn]{IEEEtran} 

\usepackage{amsmath,amssymb,amsfonts}
\usepackage[final]{graphicx}
\usepackage{psfrag}
\usepackage[numbers,sort&compress]{natbib}
\usepackage{epstopdf}
\usepackage{multirow}
\usepackage[nolist,nohyperlinks]{acronym}

\usepackage{xparse}
\NewDocumentCommand{\evalat}{sO{\big}mm}{%
  \IfBooleanTF{#1}
   {\mleft. #3 \mright|_{#4}}
   {#3#2|_{#4}}%
}

\usepackage{pgf}

\usepackage{tikz}

\usetikzlibrary{arrows,automata,shapes.geometric,shadows,positioning,babel}

\tikzstyle{line} = [draw, -latex]

\tikzstyle{noarrow} = [draw, -]

\usepackage{circuitikz}

\usepackage{graphicx}

\usepackage{stackengine}

\begin{acronym}
\acro{PLS}{physical layer security}
\acro{ASC}{average secrecy capacity}
\acro{RV}{random variable}
\acro{SNR}{signal-to-noise ratio}
\acrodefplural{SNR}[SNRs]{signal-to-noise ratios}
\acro{AWGN}{additive white Gaussian noise}
\acro{CSI}{channel state information}
\acro{PDF}{probability density function}
\acrodefplural{PDF}[PDFs]{probability density functions}
\acro{CDF}{cumulative distribution function}
\acro{cCDF}{complementary cumulative distribution function}
\acrodefplural{CDF}[CDFs]{cumulative distribution functions}
\acro{BS}{base station}
\acro{TDD}{time-division duplexing}
\acro{UL}{uplink}
\acro{DL}{downlink}
\acro{MRT}{maximal ratio transmission}
\acro{MRC}{maximal ratio combining}
\acro{MC}{Monte Carlo}
\acro{MIMO}{multiple-input multiple-output}
\end{acronym}

\makeatletter
\def\blfootnote{\xdef\@thefnmark{}\@footnotetext}
\makeatother

\begin{document}
\title{{A Product Channel Attack to\\ Wireless Physical Layer Security}}
\author{Gonzalo J. Anaya-Lopez, Gerardo Gomez and F. Javier Lopez-Martinez}

\maketitle
\blfootnote{\noindent Manuscript received July xx, 2020; revised XXX. This work has been funded by the Spanish Government and the European Fund for Regional Development FEDER (projects TEC2016-80090-C2-1-R and TEC2017-87913-R), by Junta de Andalucia (project P18-RT-3175, TETRA5G) and by University of M\'alaga. The review of this paper was coordinated by XXXX. }

\blfootnote{\noindent The authors are with Departmento de Ingenieria  de Comunicaciones, Universidad de Malaga - Campus de Excelencia Internacional Andalucia Tech., Malaga 29071, Spain  (e-mail: $\rm \{gjal, ggomez, fjlopezm\}@ic.uma.es$).}

\blfootnote{Digital Object Identifier 10.1109/XXX.2020.XXXXXXX}

\blfootnote{This work has been submitted to the IEEE for publication. Copyright may be transferred without notice, after which this version may no longer be accessible.} 

\begin{abstract}
We propose a novel attack that compromises the physical layer security of downlink (DL) communications in wireless systems. This technique is based on the transmission of a slowly-varying random symbol by the eavesdropper during its uplink transmission, so that the equivalent fading channel observed at the base station (BS) has a larger variance. Then, the BS designs the secure DL transmission under the assumption that the eavesdropper's channel experiences a larger fading severity than in reality. We show that this approach can lead the BS to transmit to Bob at a rate larger than the secrecy capacity, thus compromising the system secure operation. Our analytical results, corroborated by simulations, show that the use of multiple antennas at the BS may partially alleviate but not immunize against these type of attacks.
\end{abstract}


\section{Introduction}
During the last decade, the research on wireless security has experienced a paradigm-shift due to the advent of \ac{PLS} techniques. Even though the broadcast nature of wireless transmission could be thought to be detrimental for security, the seminal works in \cite{Bloch2008,Gopala2008} paved the way to leverage the random nature of wireless channels to provide information-theoretic security to communications in the presence of eavesdroppers. Similarly to conventional techniques for security provision in higher layers, \ac{PLS} in wireless environments is also sensitive to attacks, with the ultimate goal of precluding a secure communication between the legitimate peers Alice and Bob. 

In addition to malicious jamming \cite{Shu2013}, which affects the ability of Alice to acquire the \ac{CSI} from Bob (thus degrading the achievable secure rate), there are other approaches proposed in the literature to perform attacks from a \ac{PLS} perspective. For instance, in massive \ac{MIMO} contexts, a pilot-contamination attack can be used to influence the base station's beamforming design \cite{Kapetanovic2015}. In this strategy, usually referred to as \emph{pilot spoofing} \cite{Tugnait2018}, an eavesdropper (Eve) transmits the same pilot sequence as Bob's during the uplink training phase, in a perfectly synchronized fashion. By doing so, the equivalent channel for the eavesdropper improves as the number of antennas is increased, which causes the secrecy capacity not to be increased as the number of antennas grows.

In this work, we propose a new type of attack aimed at compromising the physical layer security in scenarios on which the potential eavesdroppers are system agents; this is the case, for instance, of wireless communication systems served by a \ac{BS}. In our scheme, the eavesdropper designs its \ac{UL} transmission in a way that the equivalent channel observed by the \ac{BS} is the product\footnote{Hence, we propose the term \emph{product channel attack} to denote this technique.} of the actual fading and a slowly-varying random sequence -- i.e, a \emph{synthetic} fading coefficient. By doing so, the \ac{BS} designs its \ac{DL} transmission to the legitimate user under the premise that the eavesdropper experiences a more severe fading than the actual one, thus choosing a secrecy rate larger than the \emph{true} secrecy capacity. The efectiveness of this technique is verified both theoretically and by simulation, showing that multiple antennas at the \ac{BS} does not suffice to improve robustness against the attack.

This strategy differs from the random beamforming or artificial fast fading scheme in \cite{Wang2015} in several aspects: (\emph{a}) our scheme is intended to operate in scenarios on which Eve is a part of the system, so that Eve's \ac{CSI} is available at Alice; (\emph{b}) we implement the artificial \emph{slow} fading technique at the eavesdropper's side, aiming to deceive Alice into assuming that Eve's channel has a larger variance; (\emph{c}) the artificial fast fading proposed in \cite{Wang2015} changes at a faster rate that the actual fading channel, so that Eve is unable to estimate the \ac{CSI}; in our case, the synthetic fading generated by Eve changes at the same rate as the actual fading, so that Alice cannot separate the effects of both sources of randomness.

The remainder of this letter is organized as follows. The system model under analysis is described in Section II, and the product channel attack technique is described in Section III. The derivation of the relevant secrecy metrics are carried out in Section IV. Performance results are evaluated and discussed in Section V. Finally, we draw the main conclusions in Section VI.

\section{System Model}
Let us assume a wireless communication system where a \ac{BS} transmits information to a set of users $\mathcal{V}$ on its coverage area. Without loss of generality, we consider that the \ac{BS} is equipped with $M$ antennas whereas users are equipped with single-antenna devices. We assume that the system operates on a \ac{TDD} mode, so that \ac{CSI} for each user can be estimated during the \ac{UL} transmission phase. We consider that all radio channels are subject to independent quasi-static Rayleigh fading, and remain constant along the transmission of each codeword.

The \ac{BS} operates in two modes for \ac{DL} transmission, referred to as \emph{standard} and \emph{secure} modes. Under \emph{standard} operation, the \ac{BS} beamforms a set of messages $z_{v}$ with $\mathbb{E}\{|z_v|^2\}=1$ and $v\in\mathcal{V}$ to each intended user through a \ac{MRT} scheme \cite{lo1999maximum}. 
Under \emph{secure} operation, the \ac{BS} wishes to establish a secure communication with a legitimate user $v_{\rm i}=\text{B}$, now assuming that a (different) illegitimate user $v_{\rm j}=\text{E}$ aims to eavesdrop the communication. 

During the \ac{UL} phase, the signal received by the \ac{BS} at the $i$-th receive antenna can be expressed as:
\begin{equation}
y_{u}^{(i)}=\sqrt{P_u L_u}x_u h_{u}^{(i)}+n^{(i)},
\end{equation}
where now $u=\{\text{B},\text{E}\}$ is used to denote the parameters corresponding to the transmission from the legitimate (B) or eavesdropper (E) users, respectively. The transmitted symbols $x_u$ are normalized so that $\mathbb{E}\{|x_u|^2\}=1$; $P_u$ represents the transmission power for user $u$; $L_u$ is the path loss measured at a reference distance $R_u$, computed as $R_u^{-\alpha}$, where $\alpha$ is the path-loss exponent; and the channel coefficients $h_{u}^{(i)}$ are circularly symmetric complex Gaussian random variables with $\mathbb{E}\{|h_{u}^{(i)}|^2\}=1$. Finally, $n^{(i)}$ represents the \ac{AWGN} samples at the $i$-th receive antenna. We assume that the \ac{BS} is able to perfectly estimate all the channel coefficients $h_u^{(i)}$, in order to recover the \ac{UL} messages, as well as to use such \ac{CSI} to design the \ac{DL} transmission.

During the \ac{DL} transmission in secure mode, the \ac{BS} beamforms the message $z_{\rm B}$ with  $\mathbb{E}\{|z_{\rm B}|^2\}=1$ through a \ac{MRT} scheme \cite{lo1999maximum}. 
In this case, the beamforming vector ${\bf w}_{\rm B}\in\mathbb{C}^{M\times1}$ is adapted to Bob's instantaneous channel; using the notation in \cite{Bjornson2014}, we have ${\bf w}_{\rm B}^{\mathcal{H}}=\frac{\bf h_{\rm B}^{\mathcal{H}}}{||{\bf h_{\rm B}}||}=\frac{[h_{\rm B}^{(1)},\ldots,h_{\rm B}^{(M)}]^{*}}{\sqrt{\sum_{i=1}^M{|h_{\rm B}^{(i)}|^2}}}$, where $^{\mathcal{H}}$ denotes the Hermitian transpose and ${\bf h_{\rm B}}\in\mathbb{C}^{M\times1}$ is the vector representation of the legitimate channel. Therefore, assuming a transmission power $P_T$ for the BS, the signal received at Bob from the BS in the \ac{DL} transmission after the \ac{MRT} processing is given by
\begin{align}
\label{eqB}
y^{\text{MRT}}_{\rm B} &= \sqrt{P_T R_{\rm B}^{-\alpha}} \underbrace{{\bf{h}_{\rm B}^{\mathcal{H}}\bf w_{\rm B}}}_{h_{\rm B}^{\rm eq}}z_{\rm B}+n_{\rm B},
\end{align}
whereas the signal received at the eavesdropper is given by
\begin{align}
\label{eqE}
y^{\text{MRT}}_{\rm E} &=  \sqrt{P_T R_{\rm E}^{-\alpha}} \underbrace{{\bf{h}_{\rm E}^{\mathcal{H}}\bf w_{\rm B}}}_{h_{\rm E}^{\rm eq}}z_{\rm B}+n_{\rm E},
\end{align}
where $n_{\rm B}$ and $n_{\rm E}$ are the \ac{AWGN} noise components at each receiver, with $\mathbb{E}\{|n_{u}|^2\}=N_0$.
In \eqref{eqB} and \eqref{eqE}, the signal arriving at each receiver is affected by an equivalent scalar channel denoted as $h_{\rm B}^{\rm eq}$ and $h_{\rm E}^{\rm eq}$, respectively. Thanks to the \ac{MRT} scheme, $|h_{\rm B}^{\rm eq}|^2$ is Gamma distributed with scale and shape parameters $M$ and $M$, respectively, whereas $|h_{\rm E}^{\rm eq}|^2$ is exponentially distributed with unitary mean \cite{Shah00}.

Hence, the instantaneous \ac{SNR}s at the legitimate and eavesdropper's sides can be expressed as
\begin{align}
\label{eqB2}
\gamma_{\rm B} &= \frac{P_T R_{\rm B}^{-\alpha}}{N_0}|h_{\rm B}^{\rm eq}|^2|z_{\rm B}|^2
\end{align}
and
\begin{align}
\label{eqE2}
\gamma_{\rm E} &=  \frac{P_T R_{\rm E}^{-\alpha}}{N_0}|h_{\rm E}^{\rm eq}|^2|z_{\rm B}|^2,
\end{align}
with average \ac{SNR}s $\overline\gamma_{\rm B}=\mathbb{E}\left\{\gamma_{\rm B}\right\}=\frac{M P_T R_{\rm B}^{-\alpha}}{N_0}=M\overline\gamma_{\rm 0}$, being $\overline\gamma_{\rm 0}$ the average \ac{SNR} in the case of a single-antenna transmitter and $\overline\gamma_{\rm E}=\mathbb{E}\left\{\gamma_{\rm E}\right\}=\frac{P_T R_{\rm E}^{-\alpha}}{N_0}$. Note that $\overline\gamma_{\rm E}$ is not influenced by the number of antennas $M$ since the beamforming vector ${\bf w}_{\rm B}$ is not adapted to Eve's instantaneous channel.

\section{The product channel attack}
As previously indicated, the \ac{BS} designs its \ac{DL} transmission using a \ac{MRT} scheme for each user, using the \ac{CSI} acquired in the \ac{UL} phase. We consider that the \ac{BS} transmit with constant power, and adapts the rate and coding schemes for each user in order to operate close to capacity (in \emph{standard} mode), or to secrecy capacity (in \emph{secure} mode). Because the \ac{BS} has perfect \ac{CSI} for every user in the system (including the eavesdropper E, or Eve), it is feasible to adapt the wiretap coding scheme to every realization of the fading channels. As indicated in \cite{Bloch2008}, any average secrecy rate below the average secrecy capacity is achievable. Hence, the secure performance is captured by the \ac{ASC} of the link between the legitimate peers (the \ac{BS}, which plays the role of Alice, and Bob) in the presence of an eavesdopper E, defined as
\begin{align}
\label{eq1}
\overline{C}_{\rm S} = \mathbb{E}\{C_{\rm S}\left(\gamma_{\rm B}, \gamma_{\rm E}\right)\},
\end{align}
\noindent
where $\gamma_{\rm B}$ and $\gamma_{\rm E}$ denote the instantaneous \acp{SNR} at Bob and Eve, respectively, and $C_{\rm S}\left( \gamma_{\rm B}, \gamma_{\rm E}\right)$ is the instantaneous secrecy capacity defined as
\begin{align}
\label{eq2}
{C}_{\rm S}( \gamma_{\rm B}, \gamma_{\rm E})\underset{\gamma_{\rm B}> \gamma_{\rm E}}{=}\log_2(1+ \gamma_{\rm B})-\log(1+ \gamma_{\rm E}).
\end{align}

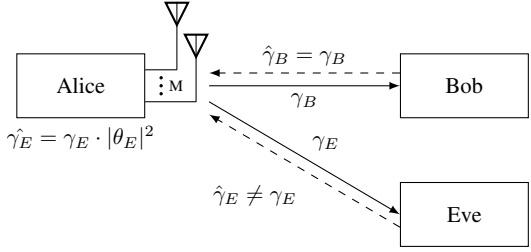
\begin{figure}
\centering
\begin{tikzpicture}[node distance = 6cm, auto, place/.style={circle,draw=black,fill=black,thick,radius=2pt}, scale=0.85, every node/.style={transform shape}]


        \node [draw, minimum width=2cm, minimum height=1cm, text centered,label=below:{$\hat{\gamma_E} = \gamma_E\cdot|\theta_{E}|^2$}] (A) {Alice};

        \node [draw, minimum width=2cm, minimum height=1cm, right of=A] (B) {Bob};

        \node [draw, minimum width=2cm, minimum height=1cm, below=1cm of B] (E) {Eve};


        \filldraw[black] ([xshift=0.25cm, yshift=0.125cm]A.east) circle (0.5pt);

        \filldraw[black] ([xshift=0.25cm]A.east) circle (0.5pt) node [right] {\scriptsize M};

        \filldraw[black] ([xshift=0.25cm, yshift=-0.125cm]A.east) circle (0.5pt);

        \path [noarrow]([yshift=0.25cm]A.east) -- ++ (0.5,0) node[antenna,scale=0.5] {};

        \path [noarrow]([yshift=-0.25cm]A.east) -- ++ (0.8,0) node[antenna,scale=0.5] {};

        \path [line]([xshift=1cm]A.east) -> node [below] {$\gamma_B$} (B);

        \path [line]([yshift=-0.25cm, xshift=1cm]A.east) -> node {$\gamma_E$} (E.west);

        \path [line]([yshift=0.2cm]B.west) [dashed] -> node [above] {$\hat{\gamma}_B = \gamma_B$} ([xshift=1cm, yshift=0.2cm]A.east);

        \path [line]([yshift=-0.2cm]E.west) [dashed] -> node [yshift=-0.1cm] {$\hat{\gamma}_E \neq \gamma_E$} ([yshift=-0.45cm, xshift=1cm]A.east);

    \end{tikzpicture}
    \caption{\ac{CSI} availability at the \ac{BS} (Alice) for the system model under consideration. For simplicity, only the users of interest are represented.}
\label{fig1}

\end{figure}

The physical layer security attack proposed in this work is formulated as follows: let us assume that during the \ac{UL} phase the eavesdropper transmits a modified symbol $\tilde{x}_{\rm E}=x_{\rm E}\cdot\theta_{\rm E}$ with $\mathbb{E}\{|\tilde{x}_{\rm E}|^2\}=1$. The synthetic variable $\theta_{\rm E}$ is generated so that it varies at the same rate as the actual fading channel, and multiplies Eve's \ac{UL} transmission symbols and pilot sequences. In this situation, the \ac{CSI} information acquired by the \ac{BS} is modified as follows: the \ac{CSI} estimated by the \ac{BS} in the \ac{UL} for Eve is now $\hat{\bf{h}}_{\rm E}=\theta_{\rm E}\bf{h}_{\rm E}$, whereas Bob's channel estimation (and hence the beamforming design) remains unaltered. Thus, the \ac{BS} is deceived into thinking that the equivalent channel observed by E after \ac{MRT} is now 
\begin{equation}
\hat{h}_{\rm E}^{\rm eq}={\hat{\bf{h}}_{\rm E}^{\mathcal{H}}\bf w_{\rm B}}=\theta_{\rm E}^{*}{{\bf{h}}_{\rm E}^{\mathcal{H}}\bf w_{\rm B}}=\theta_{\rm E}^{*}{h}_{\rm E}^{\rm eq},
\end{equation}
so that Eve's instantaneous \ac{SNR} estimation available at the \ac{BS} becomes $\hat{\gamma}_{\rm E}=|\theta_{\rm E}|^2\gamma_{\rm E}$. However, the average \ac{SNR} estimated by the \ac{BS} is not modified with respect to the case on which the attack is not performed, i.e. $\hat{\overline\gamma}_{\rm E}=\mathbb{E}\left\{\hat{\gamma}_{\rm E}\right\}={\overline\gamma}_{\rm E}$. Note that the average \ac{SNR} is chiefly determined by the path loss due to the distance between each user and the \ac{BS}, which can also be determined through timing alignment strategies. Thus, the eavesdopper avoids the risk that the \ac{BS} detects an abnormal variation in the average SNR estimated from this user. With all the above considerations, the overall \ac{CSI} availability at the \ac{BS} can be summarized as in Fig. \ref{fig1}.

Now, the transmission rate from A to B is decided by A based on the \acp{SNR} $\hat\gamma_{\rm B}$ and $\hat\gamma_{\rm E}$, and the latter differs from the actual \acp{SNR} at Eve due to the fact that the attack does not affect the legitimate channel. With the \ac{CSI} availability at A, the transmission rate towards B is then designed to maximize the following metric:
\begin{align}
\label{eqRs}
R_{\rm S}(\gamma_{\rm B},\hat \gamma_{\rm E})&\underset{\gamma_{\rm B}> \hat{\gamma}_{\rm E}}{=} C_{\rm B}(\gamma_{\rm B}) - \hat{C}_{\rm E}(\hat \gamma_{\rm E})\nonumber\\
&\underset{\gamma_{\rm B}> \hat{\gamma}_{\rm E}}{=} \log_2(1+ \gamma_{\rm B})-\log(1+ \hat{\gamma}_{\rm E})>0.
\end{align}

From the \ac{BS} perspective, the metric $R_{\rm S}$ corresponds to the secrecy capacity of the legitimate link. However, comparing \eqref{eq2} and \eqref{eqRs} it is evident that such rate does not coincide with the secrecy capacity; in the following, we will refer to $R_{\rm S}$ as \textit{compromised secrecy rate}. Using the definition in \eqref{eqRs}, the average compromised secrecy rate can be computed as
\begin{align}
\label{eqRs2}
\overline{R}_{\rm S} = \mathbb{E}\{R_{\rm S}\left(\gamma_{\rm B}, \hat{\gamma}_{\rm E}\right)\}.
\end{align}
Now, when the average \ac{SNR} at Bob is sufficiently large, the average secrecy capacity is tightly approximated by \cite{moualeu}
\begin{align}
\label{eq1b}
\overline{C}_{\rm S} \underset{\overline\gamma_{\rm B}\rightarrow\infty}{\approx}  \mathbb{E}\{C_{\rm B}\left(\gamma_{\rm B}\right)\}-\mathbb{E}\{C_{\rm E}\left(\gamma_{\rm E}\right)\},
\end{align}
and hence, we also have that
\begin{align}
\label{eq1c}
\overline{R}_{\rm S} \underset{\overline\gamma_{\rm B}\rightarrow\infty}{\approx}  \mathbb{E}\{C_{\rm B}\left(\gamma_{\rm B}\right)\}-\mathbb{E}\{\hat{C}_{\rm E}\left(\hat{\gamma}_{\rm E}\right)\}.
\end{align}
Because the equivalent channel $\hat{h}_{\rm E}^{\rm eq}$ has a larger variance that ${h}_{\rm E}^{\rm eq}$, then $\mathbb{E}\{C_{\rm E}\left(\gamma_{\rm E}\right)\}>\mathbb{E}\{\hat{C}_{\rm E}\left(\hat{\gamma}_{\rm E}\right)\}$ and hence $\overline{R}_{\rm S}>\overline{C}_{\rm S}$. Thus, when the \ac{BS} considers that E is experiencing a larger fading severity than the actual one, then A is deceived into assuming that the eavesdropper channel has a lower capacity than in reality. This causes that the \ac{BS} can select a rate $\overline R_{\rm S}$ that exceeds the secrecy capacity, which compromises physical layer security.

\section{Analytical results}
We now provide analytical expressions to give mathematical support to the performance degradation due to the proposed attack. We use the formulation of the average secrecy capacity introduced in \cite[eq. (29)]{moualeu}:
\begin{align}
\label{CS}
\overline{C}_{\rm S}({{\overline\gamma}_{\rm B}},{{\overline\gamma}_{\rm E}})=\overline C_{\rm B}(\overline\gamma_{\rm B})-\mathcal{L}\left(\overline\gamma_{\rm B},\overline\gamma_{\rm E}\right),
\end{align}
\noindent
where $\overline C_{\rm B}(\overline\gamma_{\rm B})=\mathbb{E}\{C_{\rm B}\left(\gamma_{\rm B}\right)\}$ is the average capacity of the legitimate link, and the term $\mathcal{L}\left(\overline\gamma_{\rm B},\overline\gamma_{\rm E}\right)\geq0$ can be regarded as an average secrecy capacity loss, defined as \cite[eq. (30)]{moualeu}:
\begin{align}
\label{LS}
\mathcal{L}\left(\overline\gamma_{\rm B},\overline\gamma_{\rm E}\right)\triangleq\frac{1}{\log 2}\int_0^{\infty}\frac{\overline{F}_{\rm E}(x)\overline{F}_{\rm B}(x)}{1+x}dx,
\end{align}
where $\overline{F}_{\rm B}(\cdot)$ and $\overline{F}_{\rm E}(\cdot)$ represent the \ac{cCDF} of $\gamma_{\rm B}$ and $\gamma_{\rm E}$, respectively, and $\log$ denotes the natural logarithm. As discussed in Section II, the legitimate \ac{SNR} $\gamma_{\rm B}$ is Gamma distributed with scale parameter $\overline\gamma_{\rm B}$ and shape parameter $M$:
\begin{align}
\label{cCDFGamma}
\overline{F}_{\rm B}(x)=e^{-\frac{M x}{\overline\gamma_{\rm B}}}\sum_{n=0}^{M-1}\left(\frac{M x}{\overline\gamma_{\rm B}}\right)^n \frac{1}{n!}
\end{align}
and $\gamma_{\rm E}$ is exponentially distributed with average $\overline\gamma_{\rm E}$. Using the \ac{cCDF} of the Gamma distribution, the \emph{true} secrecy capacity in \eqref{CS} can be evaluated as
\begin{equation}
\label{eqCb}
\overline C_{\rm B}\left(\overline\gamma_{\rm B}\right)=\frac{1}{\log 2}e^{\frac{M}{\overline\gamma_{\rm B}}}\sum_{n=0}^{M-1}E_{n+1}\left(\frac{M}{\overline\gamma_{\rm B}}\right),
\end{equation}
where $E_{m}(\cdot)$ is the generalized Exponential Integral, and
\begin{equation}
\label{eqLoss1}
\mathcal{L}\left(\overline\gamma_{\rm B},\overline\gamma_{\rm E}\right)=\tfrac{e^{\frac{M}{\overline\gamma_{\rm B}}+\frac{1}{\overline\gamma_{\rm E}}}}{\log 2}\sum_{n=0}^{M-1}\left(\tfrac{M}{\overline\gamma_{\rm B}}\right)^n \Gamma\left(-n,\tfrac{M}{\overline\gamma_{\rm B}}+\tfrac{1}{\overline\gamma_{\rm E}}\right),
\end{equation}
using the same procedure as in \cite{Alouini99}, where $\Gamma(\cdot,\cdot)$ is the upper incomplete Gamma function.

Analogously, the average compromised secrecy rate can be defined as:
\begin{align}
\label{RS}
\overline{R}_{\rm S}({{\gamma}_{\rm B}},{\hat{\gamma}_{\rm E}})=\overline C_{\rm B}(\overline\gamma_{\rm B})-\hat{\mathcal{L}}\left(\overline\gamma_{\rm B},\overline\gamma_{\rm E}\right)
\end{align}
\noindent
where now $\hat{\mathcal{L}}\left(\overline\gamma_{\rm B},\overline\gamma_{\rm E}\right)$ is given by
\begin{align}
\label{LS2}
\hat{\mathcal{L}}\left(\overline\gamma_{\rm B},\overline\gamma_{\rm E}\right)\triangleq\frac{1}{\log 2}\int_0^{\infty}\frac{\overline{F}_{\hat{\rm E}}(x)\overline{F}_{\rm B}(x)}{1+x}dx,
\end{align}
and $\overline{F}_{\hat{\rm E}}(\cdot)$ represents the \ac{cCDF} of $\hat\gamma_{\rm E}$. With these definitions, the condition for a successful attack (i.e., making A transmit at a larger rate than $\overline C_{\rm S}$) is given by
\begin{equation}
\label{eqD}
\mathcal{D}(\mathrm{bps/Hz})\triangleq{{\mathcal{L}}}\left(\overline\gamma_{\rm B},\overline\gamma_{\rm E}\right)-\hat{\mathcal{L}}\left(\overline\gamma_{\rm B},\overline\gamma_{\rm E}\right)>0,
\end{equation}
which is measured in excess of bps/Hz.

The distribution of $\hat{\gamma}_{\rm E}$ is that of a composite \ac{RV} built as $\hat{\gamma}_{\rm E}=|\theta_{\rm E}|^2{\gamma}_{\rm E}$, and its \ac{cCDF} can be computed as
\begin{align}
\label{CDFeq}
\overline{F}_{\hat{\rm{E}}}(z)=\int_0^{\infty}\overline{F}_{\rm E}\left(\frac{z}{y}\right)f_{\rm y}\left(y\right)dy,
\end{align}
where $f_{\rm y}\left(y\right)$ is the \ac{PDF} of the variable $y=|\theta_{\rm E}|^2$. We will now exemplify how the choice of the distribution of the synthetic symbol $\theta_{\rm E}$ impacts the physical layer security performance.
%
%


Let us first consider that $|\theta_{\rm E}|$ is drawn from a Rayleigh distribution, similar to the fading channel under consideration. Hence, the power random variable $y={\left| \theta_{\rm E} \right|^2}$ follows an exponential distribution with unitary power. The \ac{cCDF} of $\hat{\gamma_{\rm E}}$ is therefore a special case of the distribution of the product of two Gamma random variables \cite{Karagiannidis2007}:
\begin{align}
\label{cdfEveRayleigh}
\overline{F}_{\hat{\rm E}}^{\rm Ray}(z) = 2 \, \sqrt{\frac{z}{\overline{\gamma}_{\rm E}}} \, K_1 \left( \sqrt{\frac{4z}{\overline{\gamma}_{\rm E}}} \right),
\end{align}
where $K_1(\cdot)$ is the modified Bessel function of the second kind and first order. Plugging \eqref{cdfEveRayleigh} and \eqref{cCDFGamma} and the well-known expression of the Gamma distribution into \eqref{LS2} and \eqref{RS}, the average compromised secrecy rate is obtained using \eqref{eqCb} as
\begin{align}
\label{LS2b}
\hat{\mathcal{L}}^{\rm Ray}\left(\overline\gamma_{\rm B},\overline\gamma_{\rm E}\right)\triangleq\frac{1}{\log 2}\int_0^{\infty}\frac{\overline{F}^{\rm Ray}_{\hat{\rm E}}(x)\overline{F}_{\rm B}(x)}{1+x}dx,
\end{align}
which can be easily evaluated numerically with standard mathematical packages.

Taking a deeper look into the assumption of $|\theta_{\rm E}|$ to be Rayleigh distributed, it could be argued that such choice would require an arbitrary instantaneous power budget at Eve's side because of the semi-infinite support for the domain of the \ac{RV} ${|\theta_{\rm E}|}$. Hence, we also consider the case on which $|\theta_{\rm E}|$ is drawn from an uniform distribution. In this situation, the power constraint $\mathbb{E}\{|\theta_{\rm E}|^2\}=1$ is translated into a support for the \ac{RV} given by ${|\theta_{\rm E}|}\in[0,\sqrt{3}]$. Integrating the exponential distribution over the support of $|\theta_{\rm E}|$ using \eqref{CDFeq} yields
\begin{align}
\label{cdfEveUniform}
\overline{F}_{\hat{\rm E}}^{\rm Uni}(z) = e^{-\frac{z}{3\overline\gamma_{\rm E}}} -\sqrt{\frac{z\pi}{3\overline\gamma_{\rm E}}}\mathrm{erfc}\left(\sqrt{\frac{z}{3\overline\gamma_{\rm E}}}\right),
\end{align}
where $\mathrm{erfc}(\cdot)$ is the complementary error function. Hence, we can compute the \ac{ASC} loss in this case as
\begin{equation}
\hat{\mathcal{L}}^{\rm Uni}\left(\overline\gamma_{\rm B},\overline\gamma_{\rm E}\right)\triangleq\frac{1}{\log 2}\int_0^{\infty}\frac{\overline{F}^{\rm Uni}_{\hat{\rm E}}(x)\overline{F}_{\rm B}(x)}{1+x}dx,
\end{equation}
respectively. Again, this \ac{ASC} loss metric can be evaluated with accuracy using standard numerical integration routines. Alternatively, exponential-like approximations to the $\mathrm{erfc}$ function can be used to obtain approximate expressions for the \ac{ASC} loss in a similar functional form as that in \eqref{eqLoss1}.

\section{Numerical Results}
We now evaluate the performance metrics introduced in the previous section for a number of scenarios of interest. In all instances, \ac{MC} simulations have been included to double-check the validity of the analytical results.

\begin{figure}[t]
\centering
\includegraphics[width=\columnwidth]{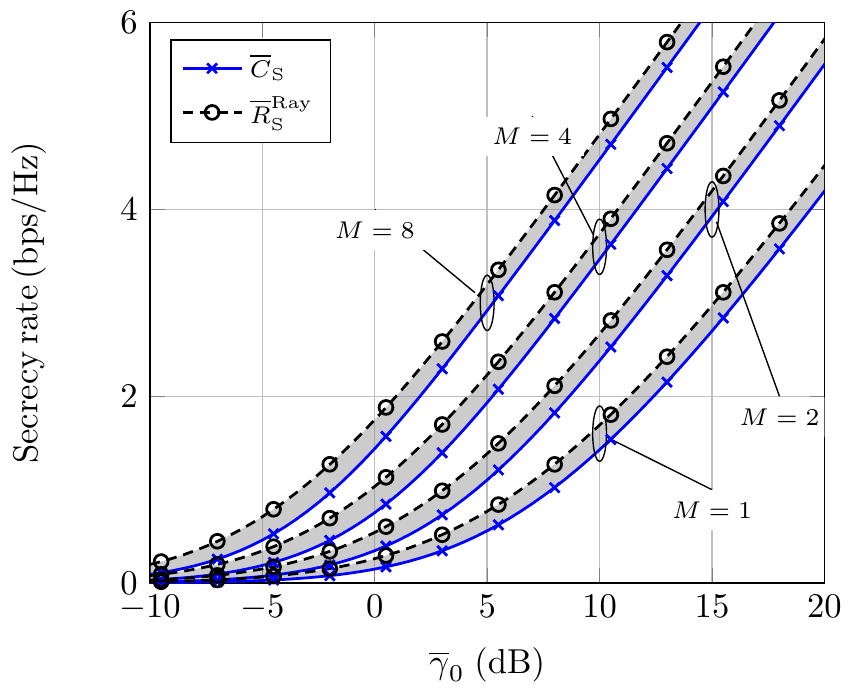} 
\caption{Average secrecy capacity $(\overline C_{\rm S})$ vs. average compromised rate $(\overline R_{\rm S})$ as a function of $\overline \gamma_{\rm 0}$, with $\overline \gamma_{\rm E}=5\;$dB and $M=1,2,4,8$. Markers correspond to \ac{MC} simulations.}
\label{f1}
\end{figure}

In Fig. \ref{f1}, the average secrecy capacity and the average compromised secrecy rate are evaluated for different antenna configurations. The eavesdropper's \ac{SNR} is set to $\overline\gamma_{\rm E}=5$dB, and the synthetic symbol $|\theta_{\rm E}|$ is drawn from a Rayleigh distribution. We observe that in all instances the compromised secrecy rate $\overline R_{\rm S}$, which is the secrecy metric available at Alice after the attack to design the \ac{DL} transmission, exceeds the \emph{true} secrecy capacity $\overline C_{\rm S}$. Hence, any transmission rate within the gray-shaded area is sensitive to be decoded by the eavesdropper.

\begin{figure}[t]
\centering
\includegraphics[width=\columnwidth]{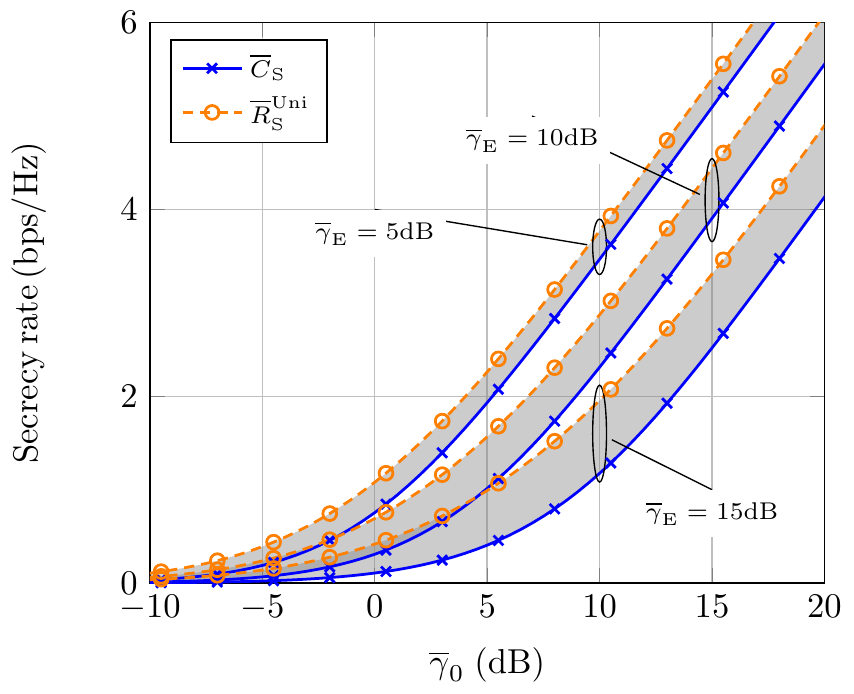} 
\caption{Average secrecy capacity $(\overline C_{\rm S})$ vs. average compromised rate $(\overline R_{\rm S})$ as a function of $\overline \gamma_{\rm 0}$, with $M=4$ and $\overline \gamma_{\rm E}=\{5,10,15\}\;$dB. Markers correspond to \ac{MC} simulations.}
\label{f2}
\end{figure}

In Fig. \ref{f2}, we now represent the average secrecy capacity and the average compromised secrecy rate for different values of $\overline\gamma_{\rm E}$. A multi-antenna transmitter with $M=4$ and a synthetic symbol $|\theta_{\rm E}|$ now drawn from a uniform distribution are considered. We see that the difference between the compromised secrecy rate $\overline R_{\rm S}$ and the \emph{true} secrecy capacity $\overline C_{\rm S}$ grows as $\overline\gamma_{\rm E}$ is increased. Hence, for a given system set-up, the attack is more harmful as Eve is closer to Alice.

\begin{figure}[t]
\centering
\includegraphics[width=\columnwidth]{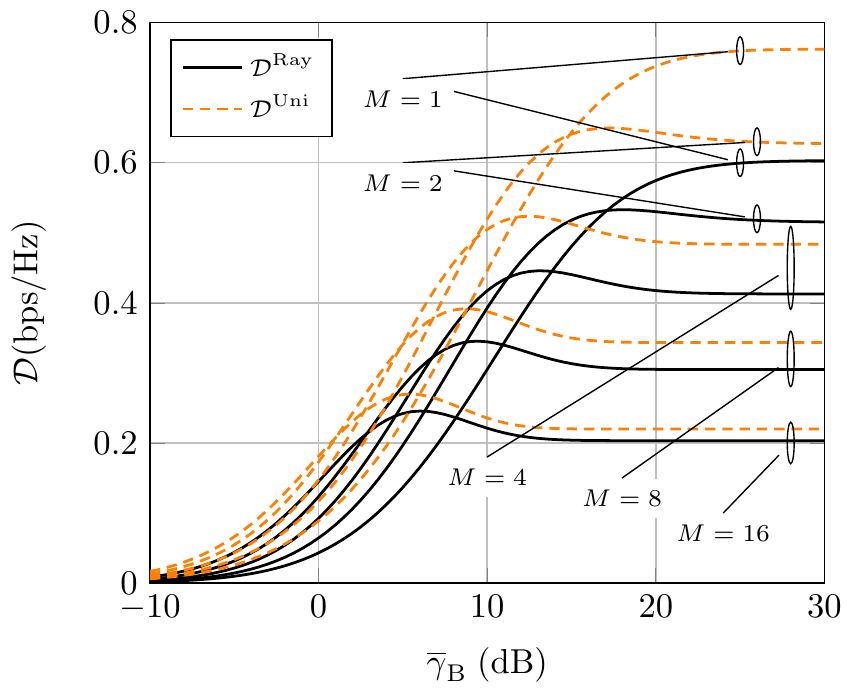} 
\caption{Excess secrecy rate $\mathcal{D}$ as a function of $\overline \gamma_{\rm B}$, for different numbers of antennas and different distributions for the synthetic symbol $\theta_{\rm E}$; $\overline \gamma_{\rm E}=15\,$dB for $M=1$ and then reduced by $10\log_{10}M\,$(dB).}
\label{f3}
\end{figure}

Finally, in Fig. \ref{f3} we evaluate the metric $\mathcal{D}$ in \eqref{eqD}, which captures the difference between the compromised secrecy rate $\overline R_{\rm S}$ and the \emph{true} secrecy capacity $\overline C_{\rm S}$. In this case, the x-axis now corresponds to the average \ac{SNR} at Bob, i.e. $\overline\gamma_{\rm B}(\mathrm dB)=\overline\gamma_{\rm 0}(\mathrm dB)+10\log_{10}M$. Hence, this means that in order to achieve a target $\overline\gamma_{\rm B}$ at Bob, the transmit power can be decreased by a factor of $M$ compared to the case of using a single antenna\footnote{We note that the scaling of $\overline{\gamma_{\rm B}}$ with $M$ does not hold in practice for arbitrarily large $M$. Hence, while it is useful to analyze the behavior of antenna arrays in practice, it should not be used for asymptotic purposes as $M\rightarrow\infty$ \cite{Bjornson2020}.}. In other words, setting a fixed value of $\overline\gamma_{\rm B}$ ideally makes $\overline\gamma_{\rm E}$ to be decreased by a factor of $M$. We can extract several important insights from the observation of Fig. \ref{f3}: (\emph i) the use of a uniformly distributed synthetic symbol seems the better choice from the perspective of the eavesdropper, as it can be generated in an easier form while at the same time allowing for a larger average compromised secrecy rate; (\emph{ii}) the use of a larger number of antennas at Alice allows for reducing $\mathcal{D}$ for high \ac{SNR} at Bob, chiefly because of the effective reduction in $\overline \gamma_{\rm E}$ for a fixed $\overline \gamma_{\rm B}$; however, as previously discussed the scaling of $\overline \gamma_{\rm B}$ with $M$ does not hold when the size of the antenna array grows \cite{Bjornson2020}. This implies that the excess secrecy rate $\mathcal{D}$ cannot be eliminated in practice by letting $M\rightarrow\infty$; (\emph{iii}) we see that in the low-\ac{SNR} regime, increasing the number of antennas at Alice is actually detrimental since the excess secrecy rate grows in this region with $M$; finally, (\emph{iv}) as $\overline\gamma_{\rm B}$ is increased the excess secrecy rate $\mathcal{D}$ saturates, so that no benefit is obtained by moving the legitimate user closer to Alice in terms of reducing $\mathcal{D}$. This behavior is well-explained by \eqref{eq1b} and \eqref{eq1c}, which make $\mathcal{D}$ to depend only on the distribution of Eve's channel, i.e.
\begin{equation}
\label{eqD2}
\mathcal{D} \underset{\overline\gamma_{\rm B}\rightarrow\infty}{\approx}\overline{C}_{\rm E}-\hat{\overline{C}}_{\rm E},
\end{equation}
where ${\overline{C}}_{\rm E}$ and $\hat{\overline{C}}_{\rm E}$ are the average capacity and the average rate estimated by Alice for  Eve's link, respectively.

With all the previous considerations, we see that for a fixed system set-up, i.e. a given number of antennas at Alice, a fixed power budget $P_{T}$ and a certain distance for the users acting as legitimate and eavesdropper agents $R_{\rm B}$ and $R_{\rm E}$, the average compromised secrecy rate will always exceed the \emph{true} secrecy capacity under this type of attack. 

\section{Conclusions}
We presented a new type of attack against wireless physical layer security, that could affect secrecy performance in scenarios where one of the system agents acts as a potential eavesdropper. The generation by the eavesdropper on the \ac{UL} phase of a synthetic symbol that varies at the same rate as the channel fading coefficients is shown to deceive the legitimate transmitter into selecting a secrecy rate that exceeds the secrecy capacity.

Since the attack becomes more effective as $\overline\gamma_{\rm E}$ grows, the use of secure areas in the proximity of Alice could help to partially mitigate the attack. Apart from this, the only choice for the \ac{BS} is to reduce the transmission rate (ideally by the same amount as the excess secrecy rate $\mathcal{D}$) so that the actual transmission rate is below the {\emph{true}} average secrecy capacity.

The proposed attack has shown to be effective even when considering eavesdroppers with the same capabilities as the legitimate agents, and without the need for using additional techniques such as jamming, multi-antenna reception or eavesdropper collusion. The impact of product channel attacks on physical layer security in more sophisticated scenarios, and the design of techniques to detect or mitigate these type of attacks seem to be interesting directions for future research activities.

\bibliographystyle{ieeetr}
\bibliography{bibfile}

\end{document}